\documentclass[journal]{IEEEtran}

\usepackage{ifpdf}
\usepackage{xcolor}

\ifCLASSINFOpdf
  
\else
  
\fi

\usepackage{graphicx}
\graphicspath{{./img/},{./img2/}}

\usepackage{amsmath}
\usepackage{amssymb}
\usepackage{multirow}
\usepackage{subfigure}

\usepackage{algorithm}
\usepackage[noend]{algpseudocode}
\usepackage{array}
\usepackage{fixltx2e}
\usepackage{stfloats}
\usepackage{tabularx}

\usepackage{lscape}
\usepackage{adjustbox}

\begin{document}


\title{ Forecasting the Spread of Covid-19 Under Control Scenarios Using LSTM and Dynamic Behavioral Models }

\author{Seid~Miad~Zandavi,
		Taha~Hossein Rashidi,
		Fatemeh~Vafaee*
\thanks{Seid M. Zandavi and Fatemeh Vafaee are with the School of Biotechnology and Biomolecular Sciences, The University of New South Wales, Sydney NSW 2052 Australia e-mails: \{s.zandavi, f.vafaee\}@unsw.edu.au.}
\thanks{Taha Hossein Rashidi is with the Research Center for Integrated Transport Innovation, School of Civil and Environmental Engineering, The University of New South Wales, Sydney NSW 2052, Australia e-mail:rashidi@unsw.edu.au.}\\
* Corresponding author}



\maketitle

\begin{abstract}

To accurately predict the regional spread of Covid-19 infection, this study proposes a novel hybrid model which combines a Long short-term memory (LSTM) artificial recurrent neural network with dynamic behavioral models. Several factors and control strategies affect the virus spread, and the uncertainty arisen from confounding variables underlying the spread of the Covid-19 infection is substantial. The proposed model considers the effect of multiple factors to enhance the accuracy in predicting the number of cases and deaths across the top ten most-affected countries and Australia. The results show that the proposed model closely replicates test data. It not only provides accurate predictions but also estimates the daily behavior of the system under uncertainty. The hybrid model outperforms the LSTM model accounting for limited available data. The parameters of the hybrid models were optimized using a genetic algorithm for each country to improve the prediction power while considering regional properties. Since the proposed model can accurately predict Covid-19 spread under consideration of containment policies, is capable of being used for policy assessment, planning and decision-making.

\end{abstract}




\section{Introduction}

\IEEEPARstart{T}{the} outbreak of coronavirus disease 2019 (Covid-19) has exposed the world to great challenges and is a serious concern for public health. The outbreak started in Wuhan, China, in December 2019 \cite{anderson2020will,kucharski2020early} and within a few weeks it spread across the globe. This caused policy changes regarding the control of the spread. There is a lack of information and uncertainty about this outbreak, making it important to understand its dynamic behavior. Forecasting the outbreak’s behavior over time can provide useful insights into the epidemiological situation \cite{camacho2015temporal} and determine whether the pandemic has been brought under control by mitigation measures \cite{riley2003transmission,funk2017impact}. Research is currently forecasting changes in infectious diseases \cite{viboud2018rapidd}, predicting the international spread of outbreaks \cite{cooper2006delaying}, and assessing the impacts of alternative interventions during pandemics \cite{kucharski2015evaluation}. 

However, research is faced with many challenges, particularly related to time. There may be delays in the presentation of symptoms due to the incubation cycle and delays in verifying detection and testing events. Delays and uncertainties can be taken into account by models, especially those stemming from normal infection histories and reporting processes \cite{nishiura2009early}. Besides, some aspects of outbreak dynamics can be biased, incomplete, or only reported by individual data sources. There is evidence that synthesis approaches will permit a more robust estimate of the dynamics underlying the transmission based on noisy data \cite{birrell2018evidence, baguelin2013assessing}. 

In order to determine the potential trajectory of the disease in accordance with the evidence, a dynamic model can be used. These predict issues such as how an infection progresses, how the number of cases/deaths is affected, or how long an outbreak lasts. In traditional approaches, SEIR models (representing the population groups susceptible, exposed, infectious and recovered) have been used to analyze the spread of Covid-19 \cite{kucharski2020early,ghaffarzadegan2020simulation}. Such models include feedback that regulates endogenous changes in contact rate, testing, diagnostics, and reporting in response to risk perception and other relevant factors. In these compartmental models \cite{kucharski2020early}, populations are divided into compartments and people are modelled by (ordinary) differential equations within the four SEIR groups.

Generative models reflect another broad variety of models with causal effects (using hidden states and parameters). For example, a generative model simulates the dynamics of effects in a group or population (i.e., new COVID-19 cases over time) \cite{friston2020dynamic}. These methods can measure the impact of policies (e.g. social distancing) and demographic variations (e.g., public immunity) in order to anticipate what might happen in a particular area under various conditions \cite{dehning2020inferring}. Accordingly, SEIR-like models have been used to imitate disease outbreaks by, for example, estimating the parameters of Bayesian Markov-Chain Monte Carlo (MCMC) models \cite{britton2002bayesian,lourenco2017epidemiological} or comprehensively discussing scenarios \cite{shulgin1998pulse,pandey2014strategies}. This family of models also has recently played a dominant role in studying the overall outbreak of the coronavirus, from inference \cite{kucharski2020early,li2020substantial} to scenario prediction \cite{anderson2020will} under control strategies \cite{zlatic2020bi}.

Here, we propose a model of time-dependent spreading movement. The time-dependence is determined through potential changes representing human and system behaviors. Therefore, a hybrid dynamic model is proposed to acquire a robust estimation of the exclusive use of region properties, which includes three main sections: A task model, facility model, and dynamic motion model. The models estimate the behavior of the Covid-19 outbreak in a particular region. The first model represents general behavior in public, such as public knowledge and how to follow the rules and knowledge. The facility model considers facilities such as hospitals, including emergency and inpatient departments, and medical staff and their knowledge, rules, and skills. The last model predicts the dynamic behavior of time-varying cases and deaths in the nominated country. The motion model uses an artificial neural network using Long Short Term Memory (LSTM) to update itself in line with stochastic behaviors and uncertainties in the proposed framework. Our framework is structured to assess past intervention efficacy and to analyze future possibilities that spread uncertainty. This model can be easily adapted to any country or region. The top-ten most-affected countries, including Australia, are studied with data collected between 31 December–19 April 2020 obtained from the European Centre for Disease Prevention and Control (ECDC). Furthermore, 90 \% of this data is used to train the hybrid dynamic model and 10 \% is used to analyze the performance of the model in the proposed framework. 

This paper is organized as follows. Section \ref{method} describes the methods. The problem formulation is explained and introduced in Section \ref{ProblemFormulation}. The results and discussion are presented in Section \ref{ResultsDiscussion}. Finally, the paper ends with a conclusion. 

\section{Method}\label{method}

Recently, deep learning has garnered the attention of many researchers in different areas. Deep learning usually defines multiple layers considering its architecture and uses a stochastic optimization algorithm to calculate the weight and bias parameters for each layer. As designed its architecture to perform the machine learning tasks, the number of depths (i.e., number of hidden layers) is directly correlated with the learning ability \cite{bengio2013representation}. In particular, LSTM, a form of recurrent neural network (RNN), is able to update during the sequence of learning as it has feedback connections, unlike a feedforward neural network. This is the key point in using LSTM to forecast with time series data. For example, it has been applied to many areas such as image captioning \cite{vinyals2015show,karpathy2015deep,mao2014deep}, natural language translation \cite{sutskever2014sequence}, and speech recognition \cite{graves2014towards}. Many of these active areas focus on classification and applications to forecast/regression models are relatively limited.

This paper aims to make a hybrid model based on LSTM and a dynamic behavioral model to forecast the spread of Covid-19 across the world from a dynamic modelling perspective. The dynamic behavioral model is introduced to describe and predict the interactions between multiple components of a phenomenon that are viewed as a system, which includes many inputs and outputs interacting over time. The dynamic model focuses on the mechanism of how the components and system evolve across time. Therefore, dynamic modeling allows us to bridge the gap between conceptualizing the phenomena of dynamic behavior and particular phenomena. Dynamic system modeling is used in many academic fields, originating in mathematics and physics before being adopted in the life, social, and behavioral sciences. It is clear that dynamic system models combined with machine learning techniques can play an essential role in data analysis and the way theories are conceived and developed.

\begin{figure*}
    \centering
    \includegraphics[scale=0.7]{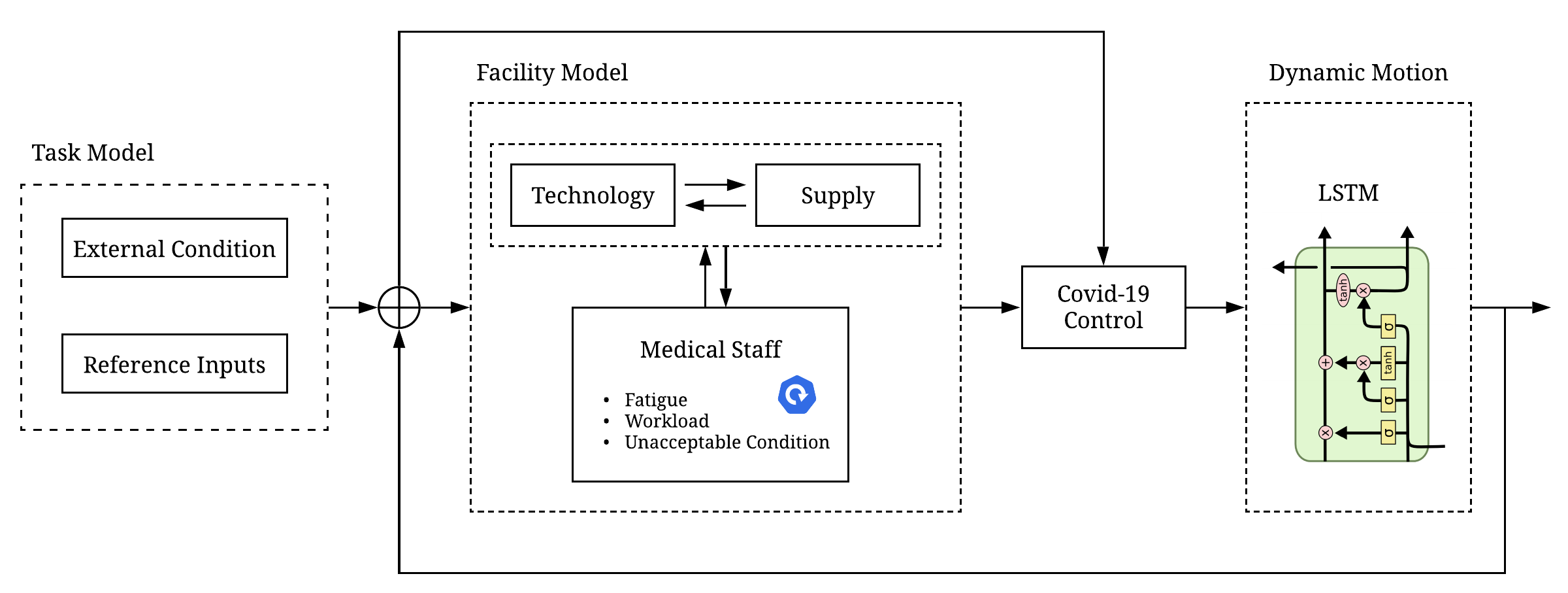}
    \caption{Hybrid Model of Covid-19 forecast}
    \label{FlowChart}
\end{figure*}

In order to forecast Covid-19, the proposed hybrid model consists of three main modules: 1) time series/sequencing learning (i.e., LSTM) with time-variant dynamics, 2) public behavior and 3) a behavioral model of the system. Figure. \ref{FlowChart} shows that Covid-19 forecasting includes three main sections. First, the task model describes external conditions and reference inputs during covid-19 epidemics. This module can describe factors such as social distancing, social knowledge, self-isolation, etc. People decide whether to conform to tasks or not, observing both external conditions and the current situation. Additionally, the reference inputs are the ideal states of public behavior during the covid-19 epidemic, so the people should control this epidemic. Second, the facility model describes covid-19 control behavior during the outbreak. This model focuses on technologies, supplies, professional personnel, etc. Such modelling can reduce the number of deaths and increase the number of discharged patients. In this model, personnel are modeled based on human features such as workload, fatigue, and conditions that affect the performance of medical staff. Lastly, dynamic motion is modelled using the artificial neural network based on LSTM to predict numbers of cases and deaths. The details of each model are introduced in the following sections. 

\subsection{LSTM Model}

LSTM is a particular form of RNN capable of learning long-term dependency, and has fundamental differences to a conventional feedforward neural network. They are sequence-based models that are able to establish the temporal correlations between previous information and current circumstances. In times series problems, like forecasting the spread of Covid-19, using a sequence-based model in an LSTM means that the decision an LSTM made at time $t-1$ affects the decision it will make at the next time, $t$. The feature (i.e., feedback connections) plays an important role in imitating the system’s dynamic motion, since it takes daily information into account when the subsequent information is entered. 

According to back-propagation through time, RNNs suffer from long-range dependencies because of gradient vanishing and exploding \cite{kong2017short}. \textit{Gradient vanishing} in RNN refers to problems where the norm of the gradient for long-term components decreases exponentially fast to zero, limiting the model’s ability to learn long-term temporal correlations, while \textit{gradient exploding} refers to the opposite event. Although LSTM has been introduced to address the issue \cite{hochreiter1997long}, the forget gate in the LSTM architecture boosts the performance of the model \cite{lipton2015critical}. This feature opens a new avenue for many sequence-learning applications. 

\begin{figure}[h]
    \centering
    \includegraphics[scale = 0.5]{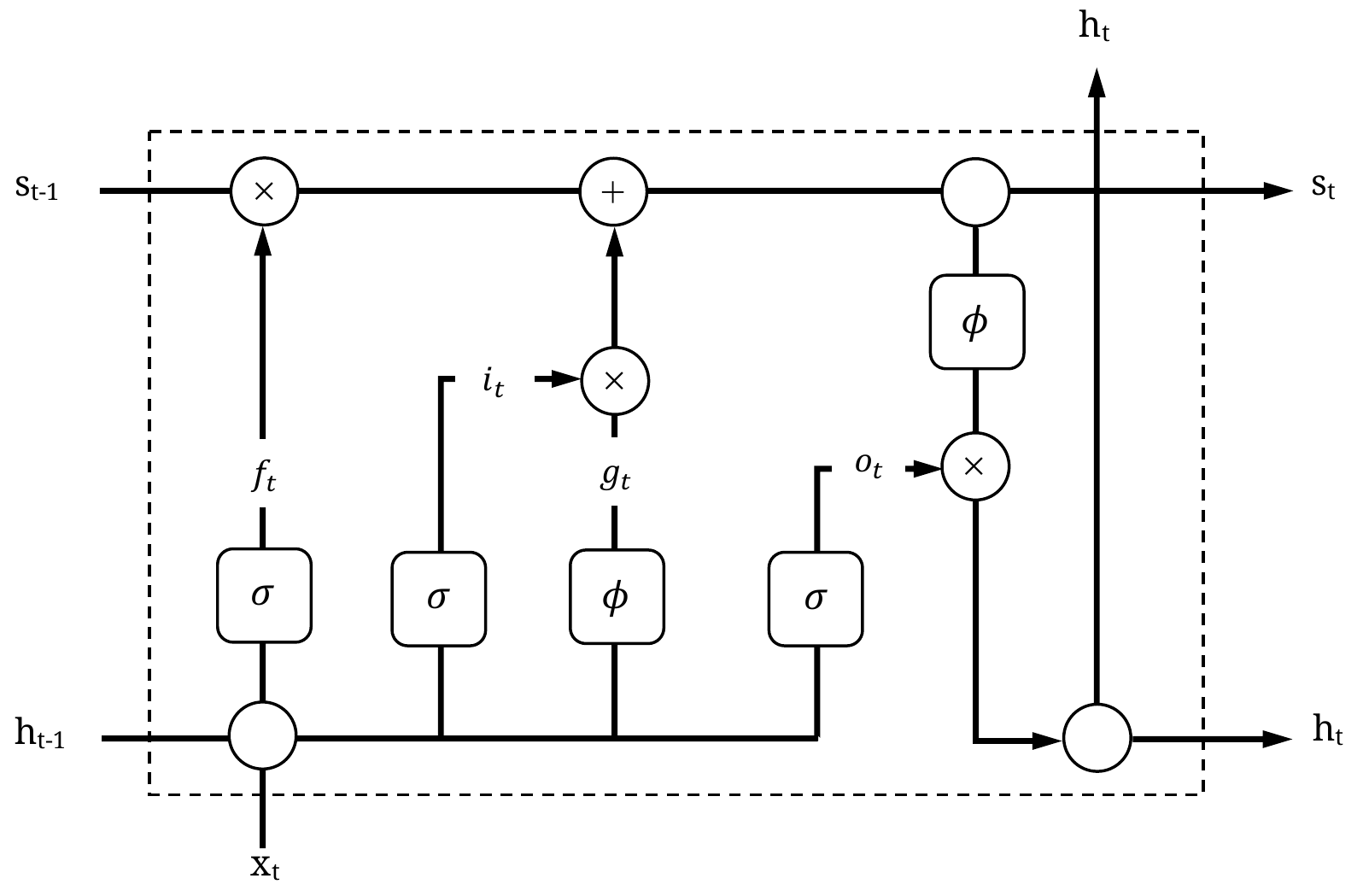}
    \caption{General Structure of LSTM}
    \label{LSTMFlowChart}
\end{figure}

Here, the general structure of LSTM is described with naming similar to that of Ref \cite{gers1999learning}. Let $\{x_1,x_2, ... , x_T\}$ present a sequencing input for an LSTM model (i.e., the general structure is illustrated in Fig. \ref{LSTMFlowChart}). In Fig. \ref{LSTMFlowChart}, $\textbf{x}_t$ is a $k$-dimensional real vector at the $t$-th time step. 

In order to establish temporal connections, the LSTM defines and maintains an internal memory cell state throughout the whole life cycle, which is the most important element of the LSTM structure. The memory cell state $s_{t-1}$ interacts with the intermediate output $h_{t-1}$ and the subsequent input $x_t$ to determine which elements of the internal state vector should be updated, maintained or erased based on the outputs of the previous time step and the inputs of the present time step. In addition to the internal state, the LSTM structure also defines an input node $g_t$, input gate $i_t$, forget gate $f_t$, and output gate $o_t$. Equation \ref{eq01} - Eq. \ref{eq06} gives the formulations for all nodes in an LSTM structure.

\begin{equation}\label{eq01}
    f_t = \sigma(W_{fx}\textbf{x}_t + W_{fh}\textbf{h}_{t-1} + \textbf{b}_f)
\end{equation}

\begin{equation}\label{eq02}
    i_t = \sigma(W_{ix}\textbf{x}_t + W_{ih}\textbf{h}_{t-1} + \textbf{b}_i)
\end{equation}

\begin{equation}\label{eq03}
    g_t = \phi(W_{gx}\textbf{x}_t + W_{gh}\textbf{h}_{t-1} + \textbf{b}_g)
\end{equation}

\begin{equation}\label{eq04}
    o_t = \sigma(W_{ox}\textbf{x}_t + W_{oh}\textbf{h}_{t-1} + \textbf{b}_o)
\end{equation}

\begin{equation}\label{eq05}
    s_t = g_t \cdot i_t + s_{t-1} \cdot f_t
\end{equation}

\begin{equation}\label{eq06}
    h_t = \phi(s_t) \cdot o_t
\end{equation}
where $W_{fx}$,$W_{fh}$, $W_{ix}$,$W_{ih}$, $W_{gx}$, $W_{gh}$, $W_{ox}$ and $W_{oh}$ are weight parameters for the corresponding input of the network activation function; $\sigma$ and $\phi$ are a sigmoid function and $\tanh(.)$, respectively. The sigmoid function with an output range of $[0,1]$ works as a soft switch for the forget gate ($f_t$), input gate ($i_t$), input node ($g_t$), and output gate ($o_t$). This means that it is a decision-making point determining whether the signal/sequencing data should pass the gate or not. For example, if the output of the sigmoid function is zero, there is no signal for the prediction. Thus, all gates (forget gate, input gates, input node and output gates), are directly depended on the current $x_t$ and previous output $H_{t-1}$.

The input gate decides what to maintain in the internal state while the forget signal is carried out from the previous state ($s_{t-1}$) by the forget gate. In order to update the internal state, the output gate points out which internal state $s_t$ should pass as the LSTM output $h_t$. This process then continues to repeat for the next time step. All the weights and biases are learned by minimizing the differences between the LSTM outputs and actual training samples. Besides, information on the current time step can be stored and maintained to affect the LSTM output of future time steps. Here, LSTM is designed to estimate the movement of Covid-19’s spread with consideration of uncertainties. The stochastic behavior of the system is modeled as a dynamic behavioral model to update the LSTM states over time.  

\subsection{Dynamic Behavioral Model}

The dynamic behavioral model describes the behavior of a system on how its elements interact with each other. The system interactions provide the functionality of the system and are used for systems and subsystems. A behavioral model represents the temporal behavior between different subsystems, while an entity interacts over time. Thus, modeling of the interactions among subsystems, the functionality of the entity over time, and the setting of roles, etc. is introduced as dynamic behavior in the system because the system performance is determined through the engagement of each module over time. 

Generally, systems and subsystems interact to accomplish a purpose by exchanging information to submit roles with expectations, from which functions are presented to ensure that action is taken. Also, an interaction over time involves communication (i.e., signals in time-series interactions), transfer of knowledge, receiving and collecting data. Thus, data manipulation/changing system behavior over time is introduced as a dynamic role. A dynamic behavioral model is constructed to demonstrate one or more interactions within the system that are responsible for task accomplishment. Accordingly, general public behavior, skilled medical staff, and high-quality hospitals cooperate as a dynamic system to manage the spread of Covid-19 in an area.


\section{Problem Formulation}\label{ProblemFormulation}
The problem is to forecast the number of cases and deaths during Covid-19 pandemics. The hybrid model using LSTM and the dynamic behavioral model are introduced to achieve good predictions. Covid-19 forecasts help to take more attention into account when the number of cases is increasing. In this section, the structure of the proposed framework (see Fig. \ref{FlowChart}) is introduced. Also, the formulation of the modules involved, such as the task model, medical staff model, hospital as facility model, and dynamic motion model, are presented in detail.


\subsection{Task Model}
In the proposed framework, the task model describes the behavior of the public in the Covid-19 time. It is categorized into two groups: an external condition and reference input. The external condition provides all the environmental factors related to people deciding whether to maintain social distancing or not. The reference input describes the ideal state in the country, such as public knowledge, keeping updated with the latest news, and government rules. It is obvious that the task model is very uncertain due to its dependency on public behavior. Thus, the uncertainties are modeled by stochastic colored noise.

The colored noise was generated by white noise that was Gaussian-distributed with a zero mean. Therefore, the dynamic is presented in Eq. \ref{eq01_1}. 

\begin{equation}\label{eq01_1}
    \frac{dx}{dt} = f(x) + \epsilon(t)
\end{equation}
where $x$ is the task model and $\epsilon$ is the white noise. Colored noise is the spectral density calculated by the Fourier-transform of the auto-correlation of white-noise ($\epsilon(t)$).  The auto-correlation is formulated in Eq. \ref{eq01_2}

\begin{equation}\label{eq01_2}
    \epsilon(t)\epsilon(k) = \frac{D}{\tau}e^{-\frac{|t-k|}{\tau}}
\end{equation}
where $D$ and $\tau$ are the noise intensity and correlation time, respectively. The Laplace transform of the auto-correlation of introduced colored noise is expressed as Eq. \ref{eq01_3}.

\begin{equation}\label{eq01_3}
    G_{Colored Noise} = \frac{A}{s+B}
\end{equation}
where $A = \frac{D}{\tau}$ and $B = \frac{1}{\tau}$. The transfer function used to model the task model is $G_{Colored Noise}$. It is obvious that the colored noise determines the uncertain behavior in terms of the physical phenomena of the motional system.


\subsection{Medical Staff Model}

According to McRuer's crossover model \cite{mcruer1974mathematical}, humans in a dynamic system behave in a way that results in an open-loop transfer function, which is formulated as a Laplace form in Eq.\ref{eq07}.

\begin{equation}\label{eq07}
    T_{op} = Y_{h} Y_{p} = \frac{\omega_c e^{-\tau s}}{s}
\end{equation}
where $Y_h$ and $Y_p$ represent human and plant transfer functions, respectively. $\omega_c$ is the crossover frequency that describes human operations and adaptation during a compensatory situation. 

\begin{figure}[!h]
    \centering
    \includegraphics[scale = 0.55]{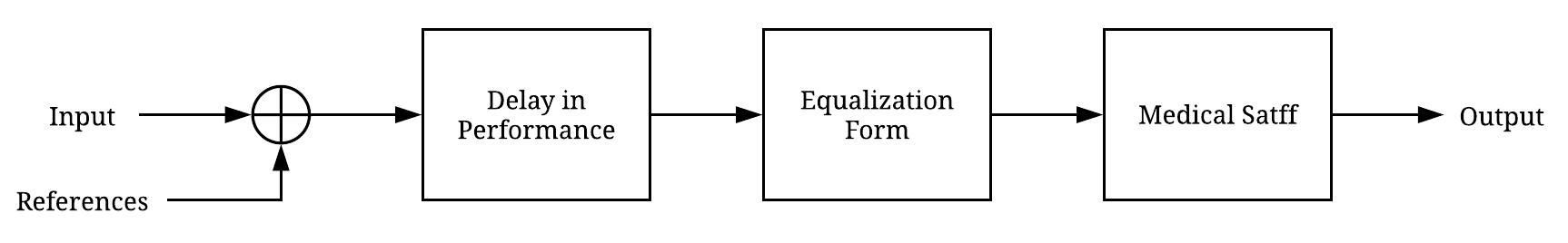}
    \caption{Human Behavior Block Diagram}
    \label{HumanModel}
\end{figure}

Figure \ref{HumanModel} shows that the human behavior model is formulated with three elements: delay in performance, equalization form, and medical staff (according to their knowledge, rules, and skills). Thus, the model is generalized and adjusted by using a describing function form and a mitigating set of rules of human characteristics, such as knowledge, rules, and skills. 

The generalized/adjusted human function, according to McRuer's crossover model, is formulated in Eq. \ref{eq08}. 

\begin{equation}\label{eq08}
    H_{staff}(s) = k_p [\frac{T_L s + 1}{T_I s + 1}] e^{-Ts} H_{KRS}(s)
\end{equation}
where $k_p$, $T_L$ and $T_I$ are the gain, time-lead and time-lag constant in the equalization from of the model shown in Fig. \ref{HumanModel}. Here, $e^{-Ts}$ represents the delay in staff response, which is described by the time-constant ($T$). $H_{KRS}$ is the transfer function of medical staff according to their knowledge, rules, and skills.  

\begin{figure}
    \centering
    \includegraphics[scale = 0.6,angle =0 ]{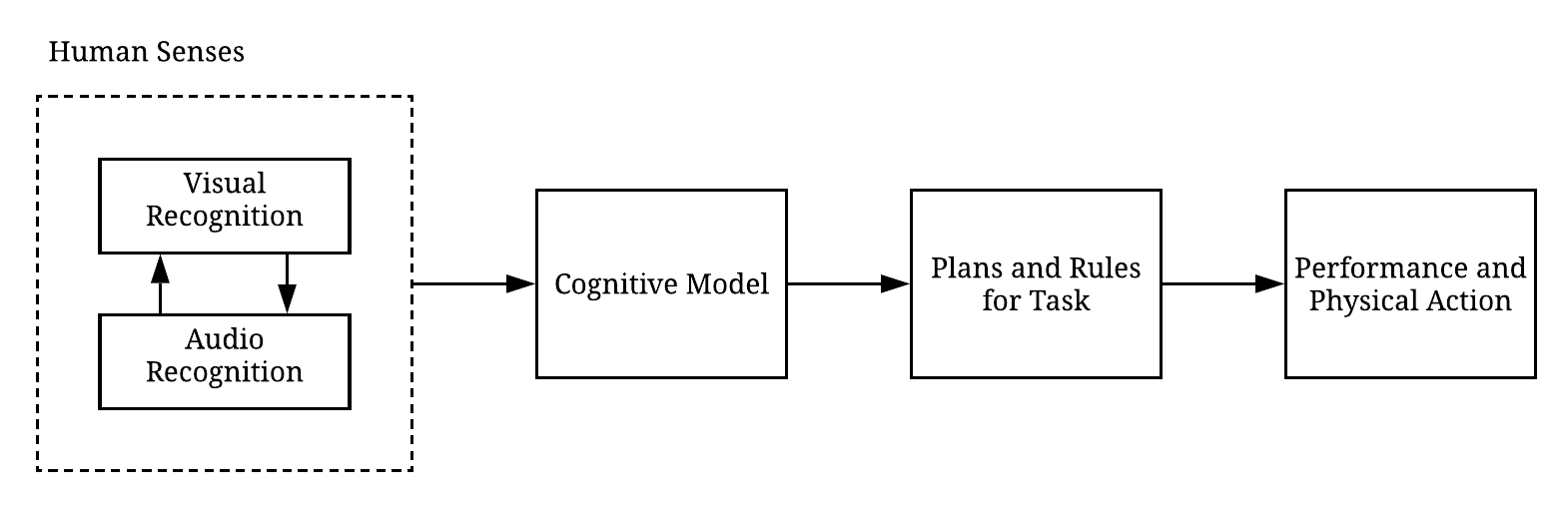}
    \caption{General Model for Knowledge, Rules and Skills}
    \label{MedicalStaff}
\end{figure}

Having generalized and adjusted the human model, knowledge, rules, and skill play important roles in staff performance. In this regard, the model consists of four modules: human senses such as visual and audio recognition, a cognitive model, plans and rules for different forms of tasks, and medical staff’s physical actions and performance efficiency. Figure \ref{MedicalStaff} illustrates the general model of medical staff. 

Human behavior and performance are highly correlated with visual and auditory recognition. These human senses are heavily dependent on workload, fatigue, and working hours. In the time-domain, the function of the human senses is modeled as an exponential function (see Eq. \ref{eq09}).

\begin{equation}\label{eq09}
    f_{hs}(t) = a+be^{-ct}
\end{equation}
where $a$, $b$ and $c$ are constant parameters in the proposed model. Laplace transform is applied to fully describe the behavior of the system. Thus, the Laplace transform of the human sense model (i.e., transfer function of human senses) is calculated as Eq. \ref{eq10}.

\begin{equation}\label{eq10}
{\begin{array}{ccc}
  F_{hs}(s) & = & \frac{a}{s} + \frac{b}{s+c}\\

   & = & \frac{(a+b)s+ac}{s(s+c)}\\
   & = & \frac{a[(\frac{a+b}{ac})s+1]}{s[\frac{1}{c}s +1]}
\end{array} } 
\end{equation}

In order to familiar form in Eq. \ref{eq10}, some replacements in the formula are made (see Eqs. \ref{eq11} - \ref{eq13} ).

\begin{equation}\label{eq11}
    k_{p_{hs}} = a
\end{equation}

\begin{equation}\label{eq12}
    T_{L_{hs}} = \frac{a+b}{ac}
\end{equation}

\begin{equation}\label{eq13}
    T_{I_{hs}} = \frac{1}{c}
\end{equation}

Alternatively, the familiar form of the human sense transfer function can be simplified as in Eq \ref{eq14}.

\begin{equation}\label{eq14}
    F_{hs}(s)= k_{p_{hs}}\frac{T_{L_{hs}}s+1}{s(T_{I_{hs}}s+1)}
\end{equation}
where $k_{p_{hs}}$ is the gain in the system, which describes the quality of human senses while submitting jobs. $T_{L_{hs}}$ and $T_{I_{hs}}$ represents the time-lead and time-lag in the proposed module. The signal output from the human sense module passes through the cognitive model, bridging the gap between understanding what to preserve and what to perform under some circumstances. As per the system defined in Fig. \ref{MedicalStaff}, an expression of the cognitive module is related to system outputs and inputs representing the conservation-of-mass principle. Besides, the cognitive model is formulated as per Eq. \ref{eq15}. 

\begin{equation}\label{eq15}
    \frac{df_{cog}(t)}{dt} = P_{cog} (Input Flow_{cog}(t)-Output Flow_{cog}(t))
\end{equation}
where $P_{cog}$ is the probability of understanding the task correctly in uncertain circumstances. The term $ Input Flow_{cog}(t)-Output Flow_{cog}(t)$ represents the length of the cognitive process in the human mind, which can be formulated as the input coming from the human senses with the delay in output considering the length of $T_{cog}$. Therefore, the formulation and transfer function of the cognitive model are as presented in Eq. \ref{eq16} and Eq. \ref{eq17}.

\begin{equation}\label{eq16}
    \frac{df_{cog}(t)}{dt} = P_{cog} (u_{cog}(t)-u_{cog}(t-T_{cog}))
\end{equation}

The Laplace transform is as below.

\begin{equation}\label{eq17}
    F_{cog}(s) = P_{cog}(1-\frac{e^{-T_{cog}s}}{s})U_{cog}(s)
\end{equation}

The transfer function is simplified in Eq. \ref{eq17_2}.

\begin{equation}\label{eq17_2}
    G_{cog} = \frac{F_{cog}(s)}{U_{cog}(s)} = P_{cog}(1-\frac{e^{-T_{cog}s}}{s})
\end{equation}

After the cognitive model has been defined, medical staff need to make a plan for their decision. Here, they need to plan and obey the rules, somehow make the limitation in jobs. In this matter, a saturation function in Laplace form is considered. 

A saturation function is introduced to define a threshold in the system’s response when the input exceeds the limit. At this time, the output becomes constant at the highest level of the threshold. Thus, the plans and rules module in the proposed system can be mathematically expressed as in Eq. \ref{eq18} .

\begin{equation}\label{eq18}
    F_{pr}(s) = \Bigg\{ 
    \begin{array}{cc}
         +K & s>s_0  \\
         s & -s_0 <s<s_0 \\
         -K & s<-s_0 \\
    \end{array}
\end{equation}

where $F_{pr}(s)$ is the transfer function of the plans and rules module represented in Fig. \ref{MedicalStaff}.

Finally, the performance and physical actions of medical staff are like a second-order dynamic system. According to the McRuer crossover theorem, the proposed model can be formulated by Eq. \ref{eq19}.

\begin{equation}\label{eq19}
    A \frac{d^2 {f}_{a}(t)}{dt^t} + B \frac{d{f}_a(t)}{dt} + c f_a(t) = d u_{pr}(t)
\end{equation}
where $A$, $B$, $C$, $D$ are constant parameters related to the nature of the system. $f(t)$ is the performance response when the plans and rules are followed by the staff. The Laplace transform of this proposed model is calculated as per Eq. \ref{eq20}. 

\begin{equation}\label{eq20}
    F_a(s) = \frac{d}{A s^2 + B s + c} U_{pr}(s)
\end{equation}

The transfer function is replaced with a similar form as that in Eq. \ref{eq21}. 

\begin{equation}\label{eq21}
    G_a(s) = \frac{F_a(s)}{U_{pr}(s)} = \frac{1}{\frac{1}{\omega_n}s^2 + \frac{2 \xi}{\omega_n}s + 1}
\end{equation}
where $\omega_n$ and $\xi$ are the natural frequency and damping ratio, respectively. These parameters define how the dynamic model can behave in the system.  

\subsection{Hospital Model}
When viewing a hospital as a system, it can be divided into two main departments—emergency and inpatients—which are the key departments that affect the quality and timeliness of patient care. Emergency capacity must be flexible throughout the day as patients arrive according to a non-homogeneous arrival pattern. Also, the inpatient sector usually focuses on maintaining bed occupancy levels to improve efficiency in terms of utilizing resources. The proposed model describes the maximum occupancy level and planned capacity. 

\begin{figure*}
    \centering
    \includegraphics[scale=0.7]{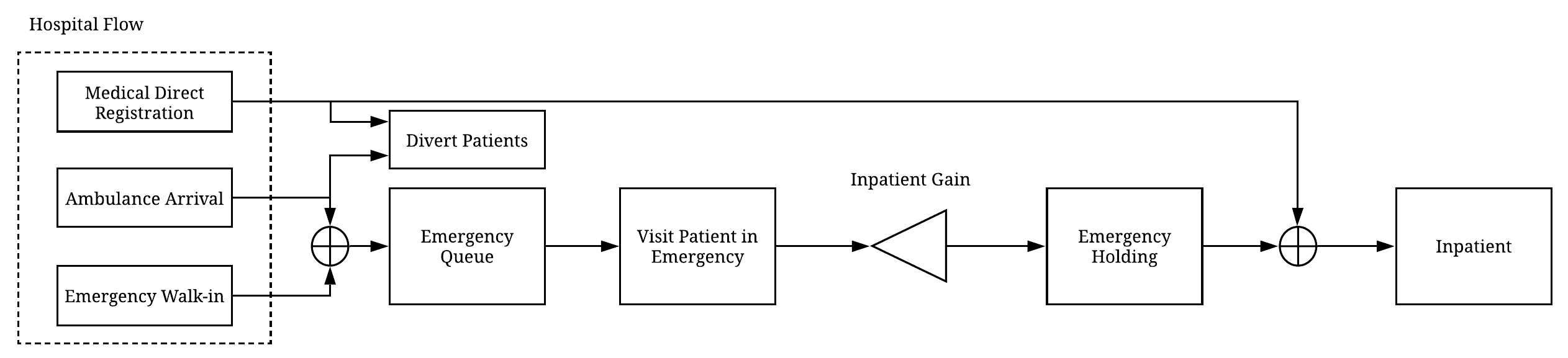}
    \caption{General Model of Hospital}
    \label{HospitalModel}
\end{figure*}

The proposed hospital model consists of three patient arrival sources: medical direct admissions, emergency walk-in patients, and emergency ambulance arrivals. While the emergency department is crowded with a considerable number of walk-in and ambulance arrival patients, diversion might carry out in the ambulance arrival. Under emergency diversion, the number of patient arrivals is reduced as ambulances are rerouted to nearby alternative hospitals. At this point, emergency walk-in patients must be registered. Besides, medical direct admission patients are sent directly to the inpatient department upon arrival to the hospital. Similar to ambulance arrivals, accepting patients to the inpatient department is heavily dependent on the availability of beds. In this regard, the diversion might be performed when crowding becomes a problem. Note that the inpatient department includes many individual sectors/units, such as the intensive care unit (ICU), telemetry units, medical/surgical units, etc. However, this work focuses on the performance of hospitals with high-demand units under conditions of a Covid-19 outbreak. Figure \ref{HospitalModel} presents the general model of a hospital during a Covid-19 pandemic with consideration of high-demand departments such as emergency and inpatient departments.

According to the conservation-of-mass principle, continuous-time patient flow in the proposed hospital model can be formulated using differential equations according to disturbances and manipulated variables to measured outputs. Equations \ref{eq22} to \ref{eq27} are formulated to describe changes and hospital tasks. Ambulance arrivals are correlated with hospital decisions of accepting or diverting some or all patients to nearby hospitals. This matter is formulated as Eq. \ref{eq22}.  

\begin{equation}\label{eq22}
    \frac{df_{h_1}}{dt} = u_1(t)
\end{equation}
where $f_{h_1}(t)$ is ambulance diversion and $u_1(t)$ describes the decision between ambulance arrival or ambulance diversion. Additionally, the emergency queue is a module that makes decisions about transferring patients emergency queue to delivering services to each individual patient. The mathematical expression is formulated in Eq. \ref{eq23}.  

\begin{equation}\label{eq23}
     \frac{df_{h_2}}{dt} = [d_1(t) + d_2(t)] -[u_1(t) + u_2(t)]
\end{equation}
where $f_{h_2}(t)$ is the emergency queue and $d_1(t)$ and $d_2(t)$ are the time-varying rates of ambulance and walk-in patient arrivals, respectively. $d_1(t)$ and $d_2(2)$ are modeled as colored noise due to a lack of information about the rate of arrivals. $u_2(t)$ is similar to $u_1(t)$ and represents the decision to transfer patients in emergency sector to visiting patients or delivering services. 

After completing emergency arrival, it is time to perform services in the emergency department. Here, the length of treatment is a crucial consideration. Equation \ref{eq24} shows the duration of progress in emergency treatment. 

\begin{equation}\label{eq24}
     \frac{df_{h_3}}{dt} = u_2(t)-u_2(t-\theta_{emg})
\end{equation}
where $f_{h_3}(t)$ and $\theta_{emg}$ are the emergency services and length of treatment, respectively. \textit{Emergency holding} is a decision about whether to hold patients in the emergency department or transfer them to inpatient beds. At this point, medical staff decide to transfer emergency patients directly to the inpatient department for further care. Equation \ref{eq25} describes the emergency holding decision.  

\begin{equation}\label{eq25}
     \frac{df_{h_4}}{dt} = k_{inp} u_2(t-\theta_{emg}) - u_3(t)
\end{equation}
where $f_{h_4}(t)$ represents emergency holding and $u_3(t)$ is the decision to transfer a patient in the holding sector to the inpatient department. $k_{inp}$ is the probability that emergency patients are admitted directly to the inpatient sector. Also, the rate of medical admission is modeled as a decision to deliver services directly to the patient (see Eq. \ref{eq26}). 

\begin{equation}\label{eq26}
     \frac{df_{h_5}}{dt} = u_4(t)
\end{equation}
where $f_{h_5}(t)$ and $u_4$ are the medical diversion and direct admission decisions, respectively. 

Finally, the hospital model describes the number of patients being treated in the inpatient department. The length of treatment in this sector is of primary importance. Therefore, controlling inpatient stays and outpatient services directly influences the number of Covid-19 cases. In this regard, the model is formulated as per Eq. \ref{eq27}.

\begin{equation}\label{eq27}
\begin{split}
     \frac{df_{h_6}}{dt} = [d_3(t)-d_3(t-\theta_{inp})]\\ +[u_3(t)-u_3(t-\theta_{inp})] \\ +[u_4(t)-u_4(t-\theta_{inp})]
\end{split}
\end{equation}
where $f_{h_6}(t)$ represent inpatient services. $d_3$ and $\theta_{inp}$ are the time-varying rates of direct patient admissions (i.e., modeled as colored noise) to the inpatient section and the length of treatment in the inpatient department, respectively. The Laplace transform of each module is represented in Eq. \ref{eq28} to Eq. \ref{eq33}. 

\begin{equation}\label{eq28}
    F_{h_1}(s) = \frac{1}{s}U_1(s)
\end{equation}

\begin{equation}\label{eq29}
     F_{h_2}(s) = \frac{1}{s} \{ D_1(s) + D_2(s) - [U_1(s)+U_2(s)] \}
\end{equation}

\begin{equation}\label{eq30}
    F_{h_3}(s) = \frac{1}{s}(1-e^{-\theta_{emg}s})U_2(s)
\end{equation}

\begin{equation}\label{eq31}
    F_{h_4}(s) = \frac{k}{s}e^{-\theta_{emg}s}U_2(s) - \frac{1}{s}U_3(s)
\end{equation}

\begin{equation}\label{eq32}
    F_{h_5}(s) = \frac{1}{s}U_4(s)
\end{equation}

\begin{equation}\label{eq33}
\begin{split}
    F_{h_6}(s) = \frac{1}{s}(1-e^{-\theta_{inp}s})D_3(s) \\ +\frac{1}{s}(1-e^{-\theta_{inp}s})U_3(s) \\
    +\frac{1}{s}(1-e^{-\theta_{inp}s})U_4(s)
\end{split}
\end{equation}


\subsection{Dynamic Motion}
\begin{figure}
    \centering
    \includegraphics[scale=0.7]{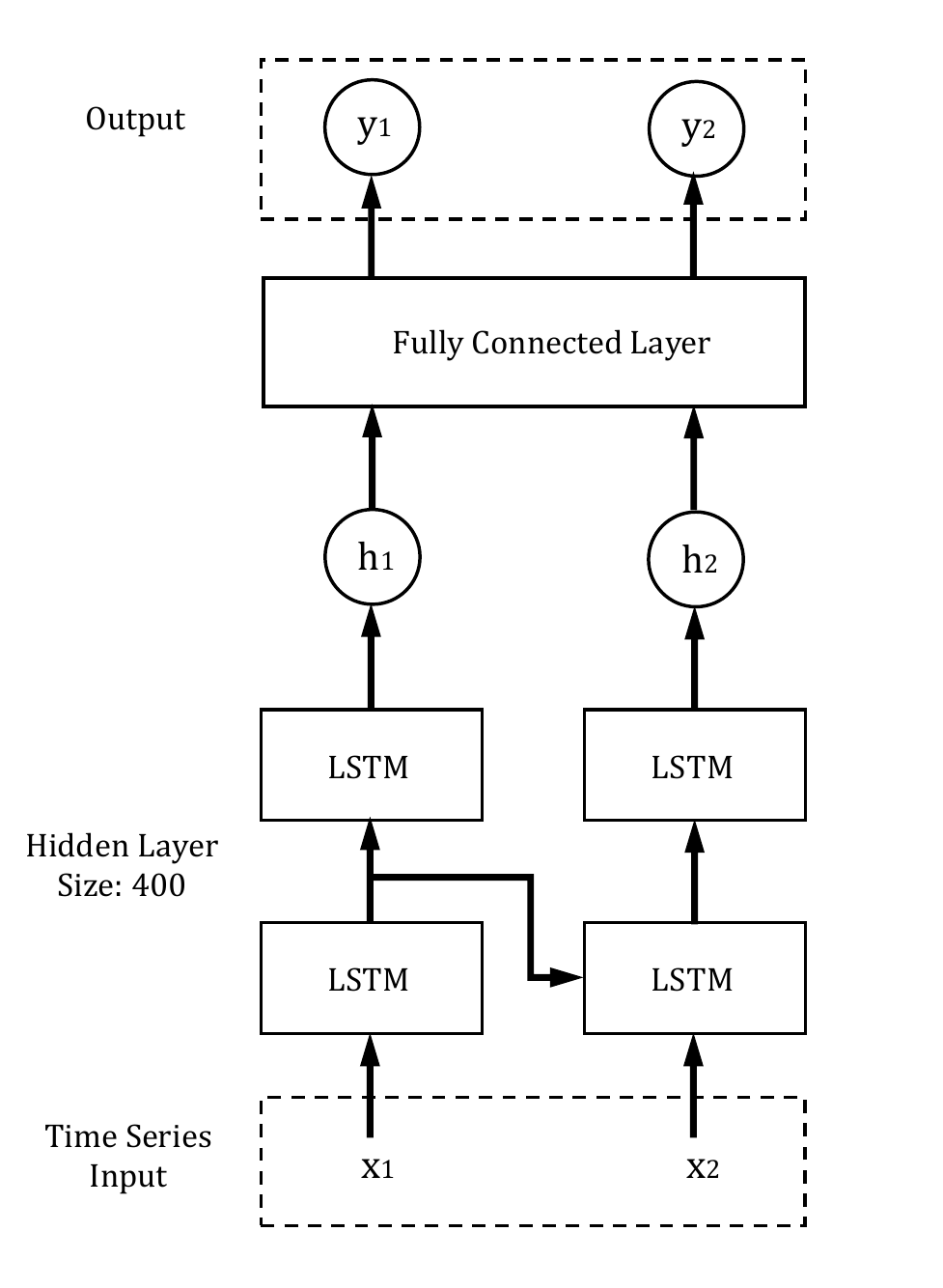}
    \caption{LSTM Architecture in Hybrid Model }
    \label{ModelLSTM}
\end{figure}

The dynamic motion model is introduced to describe time-varying cases and deaths. After complete modeling of the public tasks and facilities, such as hospitals and medical staff behavior, movement is of importance in the proposed model. The time-varying movement is obviously dependent on the proposed models such as the tasks and facilities models (see Fig. \ref{FlowChart}). The proposed framework is determined by how public tasks, facilities, and medical staff behavior affect the numbers of cases and deaths during a Covid-19 outbreak. In this dynamic motion module, LSTM is utilized as a dynamic model in the proposed system. LSTM is a time-series model, which is able to estimate the temporal correlations acquired by the simulation model and real data from previous and current circumstances. Therefore, the decision the LSTM makes at time $t-1$ affects the decision it will be make at the next time step, $t$. The LSTM's feedback connections imitate the system’s dynamic motion, since it takes daily information into account when the subsequent information is entered. Here, the proposed LSTM architecture is shown in Fig. \ref{ModelLSTM}. 

\section{Results and Discussion}\label{ResultsDiscussion}
In this section, the performance of the proposed hybrid model is evaluated using the latest available public data on Covid-19\footnote{https://www.ecdc.europa.eu/en/publications-data/download-todays-data-geographic-distribution-covid-19-cases-worldwide}. From the worldwide distribution of Covid-19 cases, the top-ten most-affected countries, which include Australia, were chosen as a case study. The regional properties used in the proposed model are fundamentally different from one another because of the different infrastructure of each region. The model properties, listed in Table \ref{tab:Parameter}, can be measured by two approaches. First, by extracting the impact of each property using a heuristic optimization algorithm \cite{zandavi2019stochastic}. Second, by utilization of actual data and the observed behavior of the whole system if it is available. In this paper, a genetic algorithm (GA) was applied to actual Covid-19 distribution data to determine the impacts of the different modules in the proposed hybrid model. Dynamic motion is represented by the LSTM, which is highly correlated with the sequence length in training the model. The sequence length plays an essential role in the prediction of Covid-19’s spread because mitigation measures and policies are being improved daily. In this regard, sequence lengths of 1–10 days are used as the primary parameters in training the LSTM. 
 
\begin{table}[h]
    \centering
    \caption{The parameters of the hybrid model}
    \begin{tabular}{clp{6cm}}
    \hline
         \textbf{Variable} && \textbf{Description} \\
         \hline
         
         $A$ && Colored noise numerator \\
         $B$ && Colored noise denominator \\
         $k_p$ && Gain in equalization \\
         $T_L$ && Time-lead in equalization \\
         $T_I$ && Time-lag in equalization \\
         $T$ && Delay time in staff response\\
         
         $k_{p_{hs}}$ && Gain in the human sense model \\
         $T_{L_{hs}}$ && Time-lead in the human sense model \\
         $T_{I_{hs}}$ && Time-lag in the human sense model \\
         $P_{cog}$ && Probability of understanding task \\
         $T_{cog}$ && Delay time in cognitive model \\
         $K$ && Saturation parameters in plans and rules model \\
         
         $\omega_n$ && Natural frequency of medical staff behavior \\
         $\xi$ && Damping ratio of medical staff behavior \\
         $\theta_{emg}$ && Length of treatment in the emergency department \\
         $\theta_{inp}$ && Length of treatment in the inpatient department \\
         $k_{inp}$ && Probability of emergency patients being transferred to the inpatient department \\
         \hline
    \end{tabular}
    
    \label{tab:Parameter}
\end{table}

\begin{figure}
    \centering
    \includegraphics[scale = 0.5]{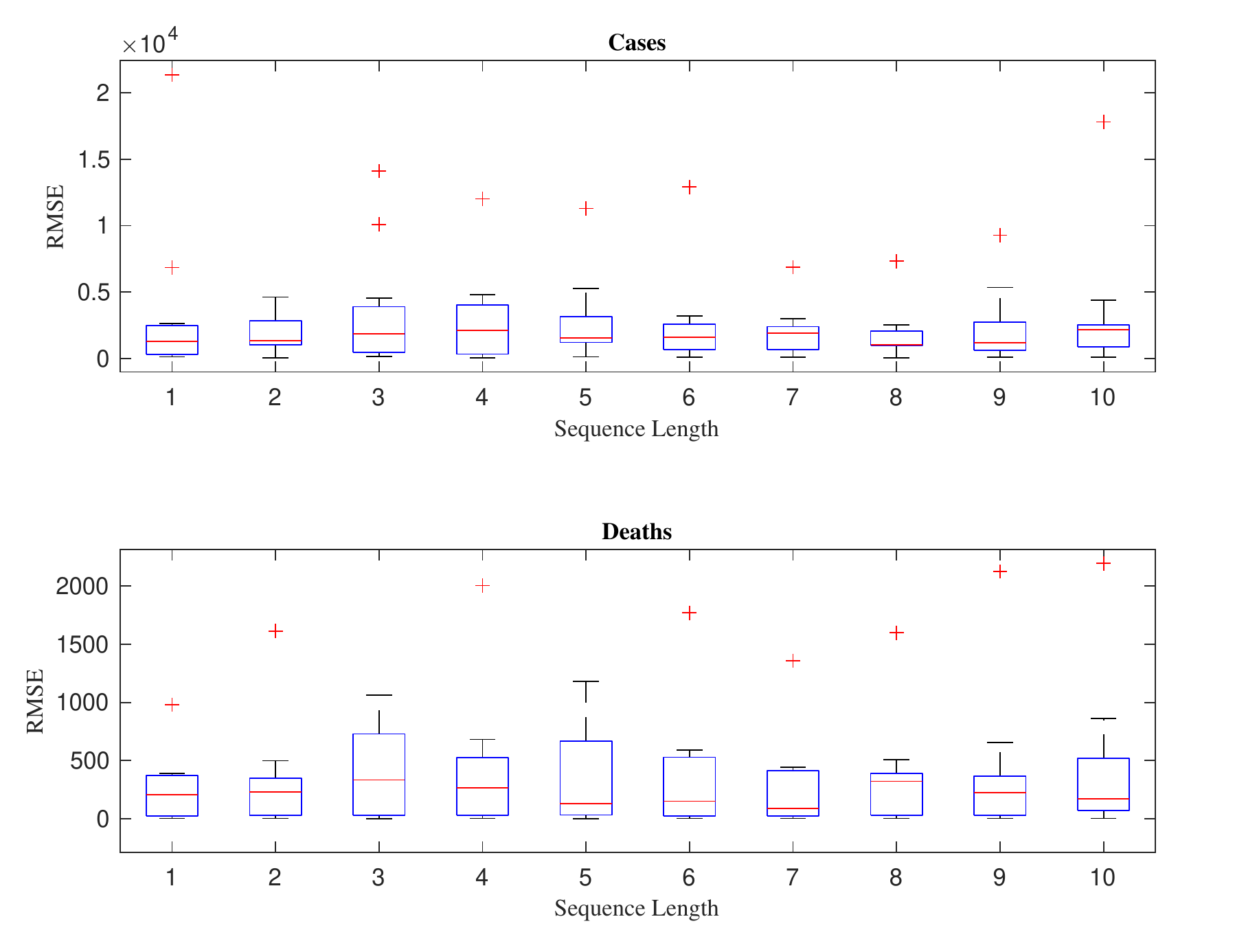}
    \caption{The Impact of Sequence Length Across the Top Ten Most-Affected Countries}
    \label{fig:LSTMSeqLen}
\end{figure}

The boxplot (see Fig. \ref{fig:LSTMSeqLen} ) shows the distribution of RMSEs across the different countries for each sequence length. A seven-day sequence length was found to provide lower variability and shorter outliers than other sequence lengths, while both cases and deaths are matter. Thus, a seven-day sequence length was selected to analyze the impacts of the different modules in the proposed hybrid model. 

Table \ref{tab:ImpactModule} lists 10 different combinations, from pure LSTM to the whole hybrid model. Generally, the proposed framework includes three main models: the Task Model (TM), Facilities Model (FM), and Dynamic Motion Model (DM). The TM includes external conditions and reference input, while the FM consists of two main sections: medical staff and hospital models. The medical staff model considers human senses, a cognitive model, plans and rules, performance and physical actions, and the hospital, which is modeled in terms of direct medical registration, ambulance arrivals, emergency walk-ins, and emergency services. The DM represents dynamic movement over time as modelled by LSTM. Here, the USA, the most-affected country, was chosen to investigate the importance of the modules in the hybrid model. 

Genetic Algorithm (GA) was used to find best parameters for each model. Accordingly, GAs with $200$ population size and $500$ performed to minimize the root mean square error (RMSE) between the predicted and observed data. The RMSE is calculated as follows: 
\begin{equation}\label{eq34}
    RMSE = \sqrt{\frac{\sum_{t =1}^{n}(\hat{x}_t-x_t)^2}{n}}
\end{equation}
where $\hat{x}_t$ and $x_t$ are the observed and predicted values, respectively. 

\begin{figure}
    \centering
    \includegraphics[scale = 0.5]{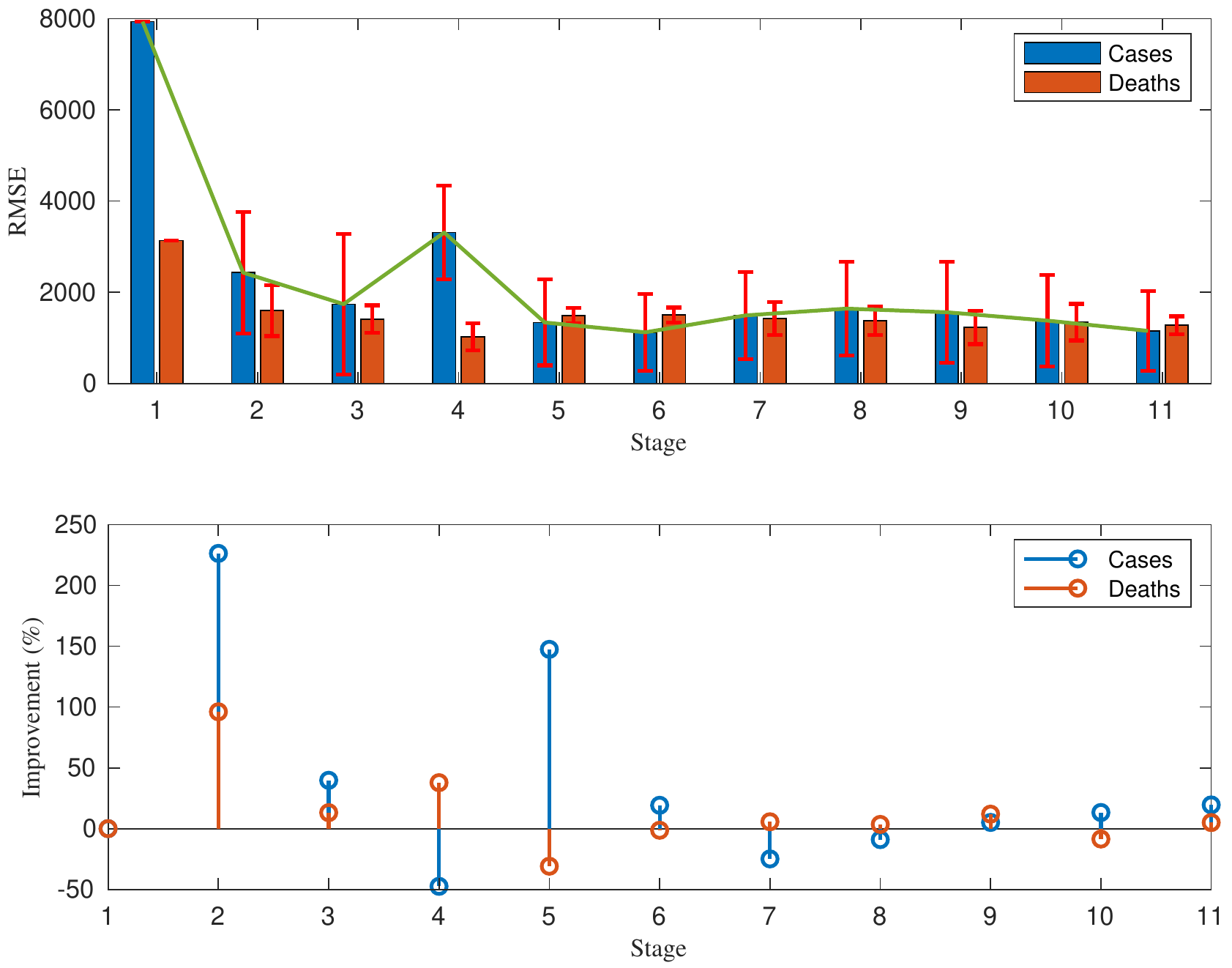}
    \caption{Cumulative Module Effect}
    \label{fig:ModuleImprovement}
\end{figure}

To assess the statistical significance of the reduction in RMSE in hybrid models compared to LSTM, each module was evaluated 500 times after hype-parameter tuning, and the corresponding RMSE distribution was used to estimate $95\%$ confidence interval (CI) and t-test p-values comparing significant differences between stage 11 and other stages. Table \ref{tab:ImpactModule} and Fig. \ref{fig:ModuleImprovement}) clearly show the cumulative effect. As seen in Fig. \ref{fig:LSTMSeqLen}, adding TM (stage 2) significantly improved the accuracy (i.e., $250\%$ better than LSTM). Then, adding more modules (i.e., greater similarity to the real environment) on top of TM (i.e., stages 3 to 11) was able to improve the model’s performance in comparison with the previous stage. 

\begin{table*}[h]
    \centering
    \caption{Cumulative effect of modules in the accuracy, estimation of $95\%$ CI and t-test p-values to compare significant differences between stage 11 and other stages }
    \begin{adjustbox}{width=\linewidth,center}
    \begin{tabular}{llp{5cm}ccccccccc}
    \hline
         \multirow{3}{*}{\textbf{Stage}} &\multirow{3}{*}{\textbf{Model}}& \multirow{3}{*}{\textbf{Module}} & \multicolumn{9}{c}{\textbf{RMSE}} \\
         
         \cline{4-12}
         
         & & & \multicolumn{4}{c}{\textbf{Case}} & &\multicolumn{4}{c}{\textbf{Death}} \\ 
         \cline{4-7}
         \cline{9-12}
         & & & \textbf{Mean} & \textbf{Std}  & \textbf{95\% CI}  & \textbf{p-value} && \textbf{Mean} & \textbf{Std}  & \textbf{95\% CI} & \textbf{p-value} \\
         \hline
         
         1 & DM & LSTM & $7934$ & $0.0$  & $(7934,7934)$ & $0.0$ && $3135$ & $0.0$  & $(3135,3135)$& $0.0$ \\
         
         2 & TM, DM & External Condition, Reference Input, LSTM & $2431$ & $1332$  &$(2374,2606)$ & $0.0$ && $1598$ & $560$  & $(1515,1612)$ & $0.0028$\\
         
         3 & TM, FM, DM & TM, Human Sense, LSTM & $1740$ & $1540$  & $(1507,2157)$ & $0.0$ && $1413$ & $300$ & $(1375,1431)$ & $1.43e-06$ \\
         
         4 & TM, FM, DM & TM, Human Sense, Cognitive Model, LSTM & $3310$ & $1024$  & $(3240,3419)$ & $0.0$&&  $1025$ & $297$  & $(998,1050)$ &  $0.0$\\
         
         5 & TM, FM, DM & TM, Human Sense, Cognitive Model, Plans and Rules, LSTM & $1337$ & $942$  &$(1257,1420)$ & $1.14e-04$ && $1483$ & $173$  & $(1470,1500)$ & $0.9340$ \\
         
         6 & TM, FM, DM &  TM, Human Sense, Cognitive Model, Plans and Rules, Human Performance, LSTM & $1122$ & $839$  & $(1040,1183)$ &  $0.6987$ && $1503$ & $161$  & $(1492,1519)$ & $0.0530$ \\
         
         7 & TM, FM, DM &  TM, Medical Staff, Ambulance Arrivals, LSTM & $1494$ & $956$ &  $(1302,1501)$ & $5.36e-06$ &&  $1423$ & $357$  & $(1396,1456)$& $0.0011$\\
         
         8 & TM, FM, DM &  TM, Medical Staff, Ambulance Arrivals, Emergency Walk-in, LSTM & $1642$ & $1029$  & $(1557,1731)$ & $6.87e-17$ && $1375$ & $312$  & $(1377,1394)$ & $1.91e-11$\\
         
         9 & TM, FM, DM & TM, Medical Staff, Ambulance Arrivals, Emergency Walk-in, Medical Direct Registration, LSTM & $1562$ & $1105$ & $(1399,1704)$ & $0.0$  && $1229$ & $365$ & $(1194,1260)$ & $0.0$\\
         
         10 & TM, FM,DM & TM, Medical Staff, Ambulance Arrivals, Emergency Walk-in, Medical Direct Registration, Emergency Services, LSTM & $ 1379$ & $1007$ & $(1362,1636)$ & $0.0$ &&  $1343$ & $401$ & $(1277,1348)$ & $8.08e-16$ \\
         
         11 & TM, FM,DM &  TM, FM (Medical Staff, Hospital), LSTM & $1154$ & $874$ & $(1057,1207)$ & - && $1279$ & $195$ & $(1267,1401)$ & - \\
         
         \hline
    \end{tabular}
    \end{adjustbox}
    
    \label{tab:ImpactModule}
\end{table*}

\begin{figure}
    \centering
    \includegraphics[scale = 0.5]{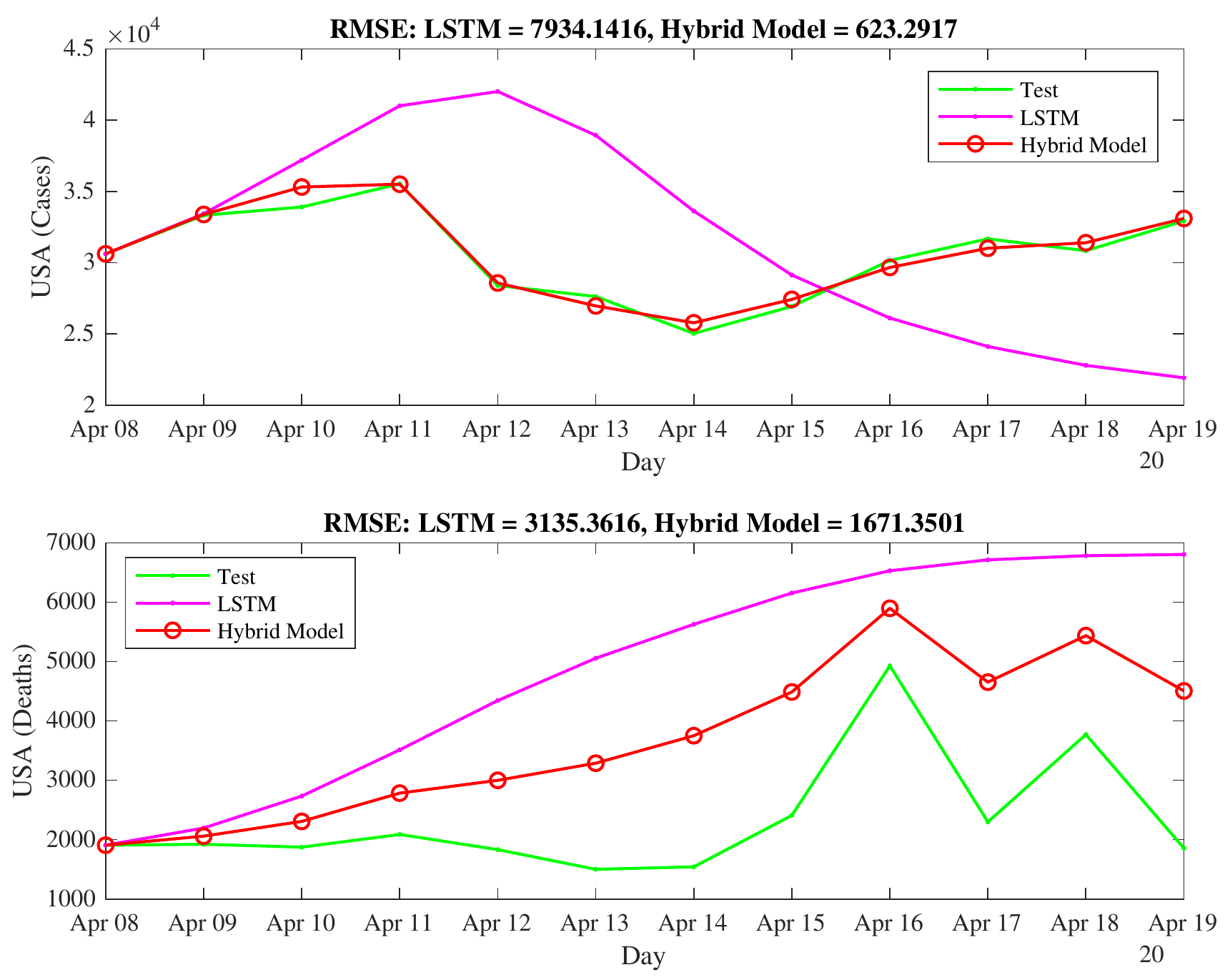}
    \caption{Performance of Hybrid Model for USA}
    \label{fig:USATest}
\end{figure}

The cumulative results show that having greater similarity to the real environment helps to obtain a more accurate model. Figure \ref{fig:ModuleImprovement} demonstrates that not only can the whole modeled system, encompassing everything from public behavior to hospital performance, reach significant accuracy, it also imitates  changes in Covid-19 time. For instance, Fig. \ref{fig:USATest} shows a comparison of LSTM and the proposed model, from which it presents that the hybrid model can more accurately estimate real ambient behavior.

\begin{figure*}
    \centering
    \includegraphics{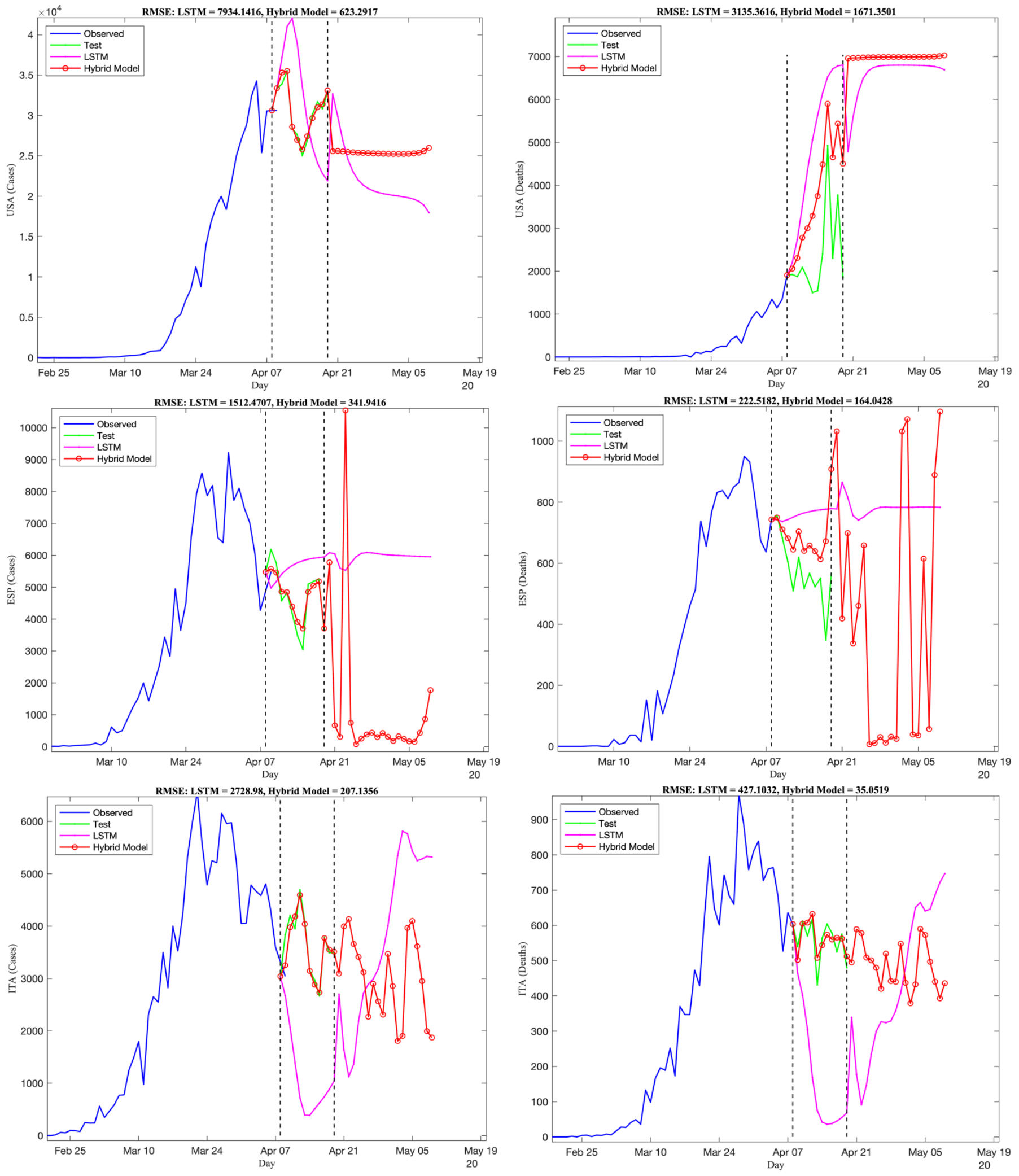}
    \caption{LSTM and Hybrid Model for USA, ESP and ITA}
    \label{fig:pic1}
\end{figure*}

\begin{figure*}
    \centering
    \includegraphics{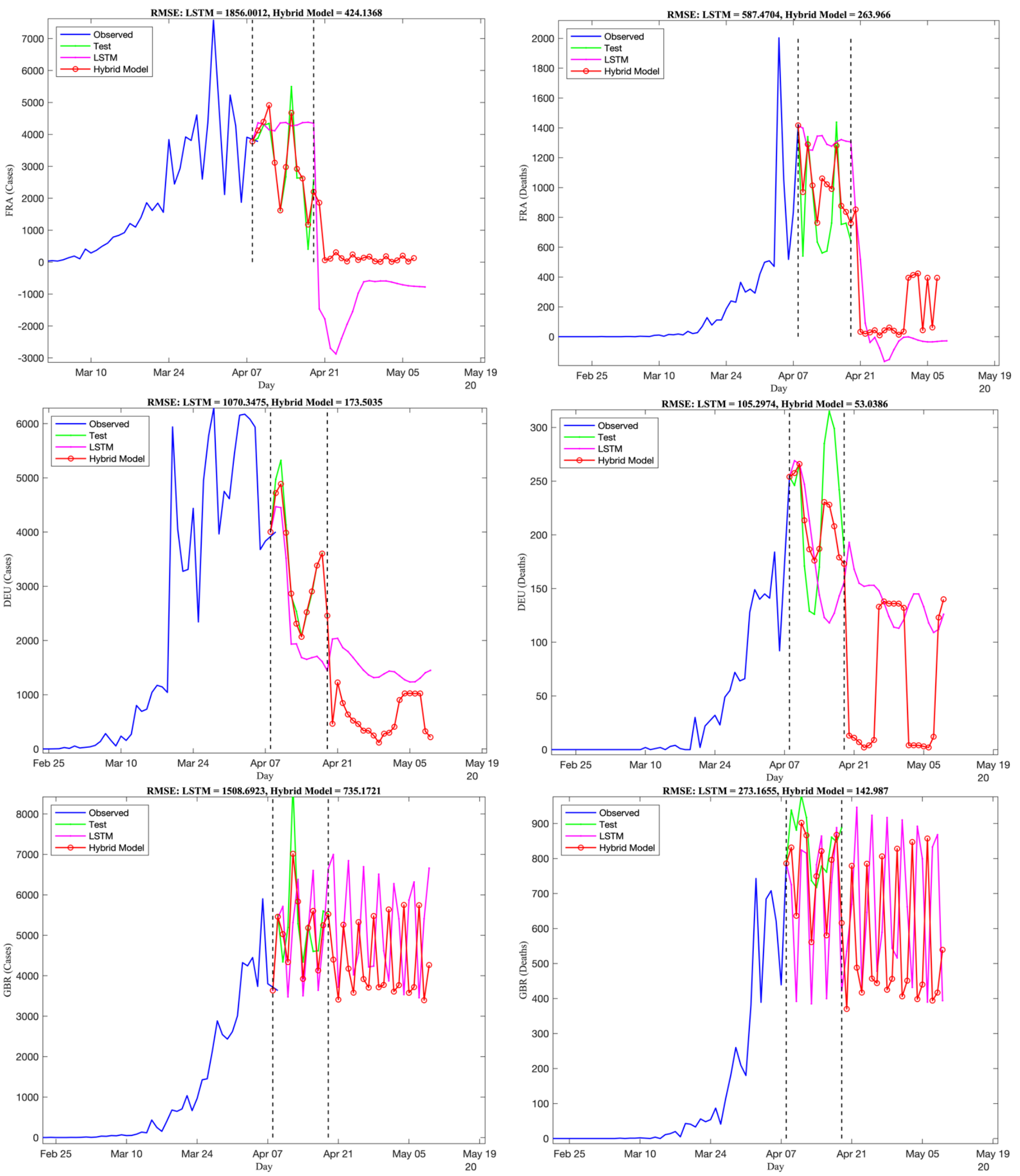}
    \caption{LSTM and Hybrid Model for FRA, DEU and BGR}
    \label{fig:pic2}
\end{figure*}

\begin{figure*}
    \centering
    \includegraphics{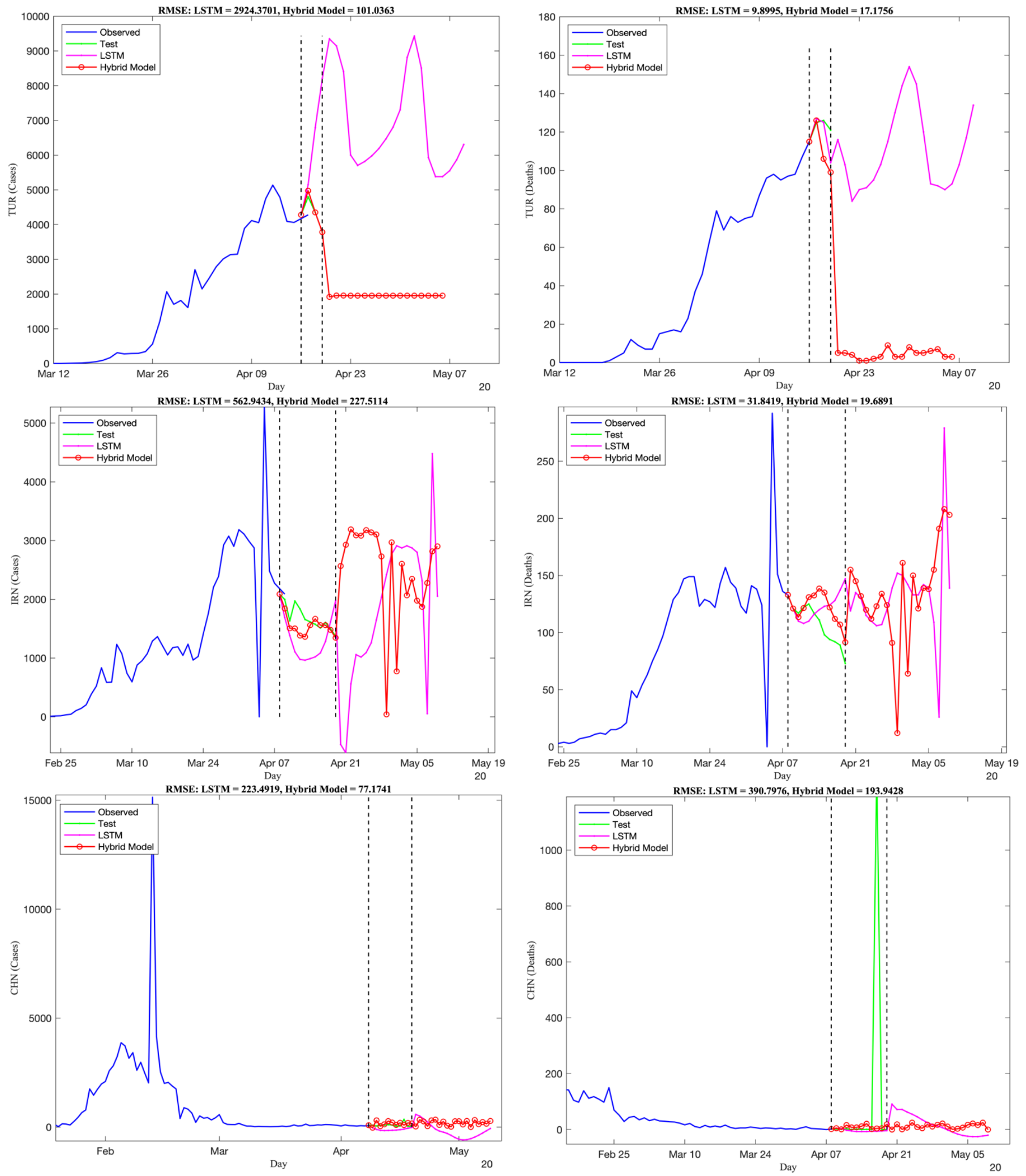}
    \caption{LSTM and Hybrid Model for TUR, IRN and CHN}
    \label{fig:pic3}
\end{figure*}

\begin{figure*}
    \centering
    \includegraphics{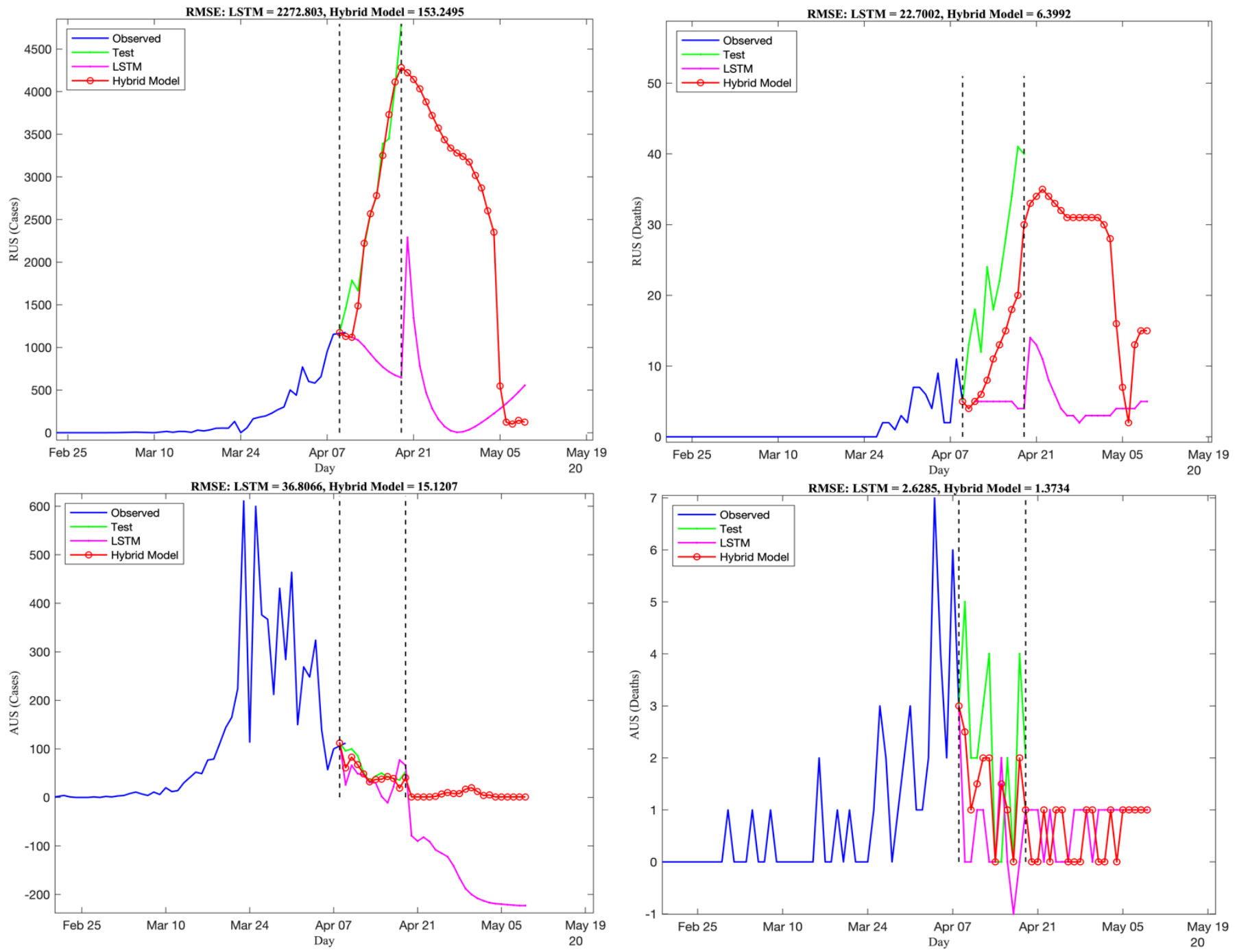}
    \caption{LSTM and Hybrid Model for RUS and AUS}
    \label{fig:pic4}
\end{figure*}

The parameters of the hybrid model were tuned by GA for the most-affected countries, which were the United States of America (USA), Spain (ESP), Italy (ITA), France (FRA), Germany (DEU), United Kingdom (GBR), Turkey (TUR), Iran (IRN), China (CHN), Russia (RUS), and Australia (AUS). The tuned parameters are listed in Table \ref{tab:table02}.
The results show that the hybrid model outperforms the LSTM model. The performance of the hybrid model in testing and forecasting is presented in Figs. \ref{fig:pic1} - \ref{fig:pic4}.


\begin{table*}
    \centering
    \caption{The tuned parameters of the hybrid model}
    \begin{adjustbox}{width=\linewidth,center}
    \begin{tabular}{cccccccccccc}
    \hline
         \textbf{Parameters} & \textbf{USA} & \textbf{ESP} & \textbf{ITA} & \textbf{FRA} & \textbf{DEU} & \textbf{GBR} & \textbf{TUR} & \textbf{IRN} & \textbf{CHN} & \textbf{RUS} & \textbf{AUS} \\
         \hline
         
         $A$            & $-0.8101$ & $-4.0729$ & $-3.3868$ & $-5.8566$ & $-2.9476$ & $-4.2758$ & $3.4529$ & $-0.7062$ & $3.5021$  & $2.0335$  & $-0.3415$\\
         $B$            & $-0.5035$ & $3.8714$  & $2.1760$  & $1.5111$  & $4.3153$  & $-1.0266$ &$-3.4325$ & $2.2764$  & $3.2667$  & $1.8693$  & $1.2992$\\
         $k_p$          & $-2.3547$ & $2.0928$  & $-2.9657$ & $-2.5787$ & $4.2061$  & $8.5490$  &$-1.9478$ & $-2.3545$ & $0.8305$  & $-0.6830$ & $1.8628$ \\
         $T_L$          & $2.5264$  & $0.9372$  & $-2.0098$ & $4.1430$  & $-2.8764$ & $0.6761$  & $4.0391$ & $1.7835$  & $-4.1396$ & $-1.8726$ & $-1.2981$ \\
         $T_I$          & $0.8912$  & $-5.4006$ & $0.8341$  & $1.0686$  & $1.4779$  & $8.8155$  &$-1.5356$ & $-3.1597$ & $1.3461$  & $3.3400$  & $1.1695$ \\
         $T$            & $1.6098$  & $1.6500$ & $0.4342 $  & $0.8359$  & $1.980$   & $10.8410$ & $1.4949$ & $1.6416$  & $6.4313$  & $1.8808$  & $5.9230$ \\
         $k_{p_{hs}}$  & $4.2978$  & $-3.5841$ & $-2.3695$ & $-0.1626$ & $-1.6574$ & $6.8063$  & $1.8011$ & $-2.2024$ & $-2.1579$ &  $3.8306$ & $2.8875$\\
         $T_{L_{hs}}$   & $-3.2900$ & $-1.7763$ & $2.7730$  & $-2.9302$ & $-2.1169$ & $-0.2613$ &$-1.3817$ & $-2.0913$ & $0.2146$  & $-1.2550$ & $-6.4650$\\
         $T_{I_{hs}}$   & $0.5158$  & $-0.1012$ & $-1.0059$ & $4.0982$  & $-0.2475$ & $-7.4055$ &$-1.5497$ & $5.6994$  & $-0.9240$ & $0.5457$  & $0.1841$ \\
         $P_{cog}$      & $0.4272$  &$0.4576$   & $0.1702$  & $0.5062$  & $0.1508$  & $0.2494$  & $0.1504$ & $0.4049$  & $0.4150$  & $0.1231$  & $0.3320$ \\
         $T_{cog}$      & $0.3461$  & $5.1084$ & $0.4207$   & $1.6048$ & $1.8097$   & $0.6702$  & $2.6286$ & $1.1620$  & $4.4143$  & $2.4355$  & $1.4879$ \\
         $K$            & $3.1723$  & $2.0726$  & $1.8494$  & $3.4353$  & $1.2141$ & $2.0622$ & $3.2185$& $1.1833$  & $0.1048$ & $1.8516$  & $2.3600$ \\
         $\omega_n$     & $3.3560$  & $4.1003$  & $1.7364$ & $2.5509$  & $0.4818$ & $2.6556$  & $0.2992$&  $2.0260$ &  $0.9775$ &  $1.9706$ & $2.1473$ \\
         $\xi$          & $1.4909$  & $2.0292$  & $2.3595$ & $1.7027$  & $0.9775$ & $1.1851$  & $0.0713$ & $2.0415$ & $2.8391$ & $1.4002$ & $0.0388$ \\
         $\theta_{emg}$ & $5.1623$  & $4.8020$  & $2.4513$  & $4.0678$  & $0.6537$  & $3.9345$  & $3.1381$ & $3.6335$  & $3.5635$  & $4.1859$  & $2.6156$ \\
         $\theta_{inp}$ & $6.6954$  & $3.7435$  & $1.9794$  & $6.9620$  & $6.9501$  & $1.0285$  & $0.6522$ & $2.9777$  & $8.4738$  & $5.6019$  & $6.2142$ \\
         $k_{inp}$      & $0.2254$ & $0.5349$  & $0.1146$  & $0.1844$  & $0.5675$  & $0.1482$  & $0.4656$ & $0.8178$  & $0.3856$  & $0.7830$  & $0.3006$\\
         
         \hline
    \end{tabular}
    \end{adjustbox}
    
    \label{tab:table02}
\end{table*}


Having tuned the hybrid model's parameters, the hybrid model was evaluated using test data. In Table \ref{tab:Comparison}, the performance of the hybrid model is compared with that of LSTM. In this table, RMSEs were obtained for cases and deaths in different countries. The results show that the performance of the hybrid model is significantly better than that of LSTM. 

\begin{table}[t]
    \centering
    \caption{Comparison of Hybrid model with LSTM}
    \begin{tabular}{cccccc}
    \hline
    \multirow{3}{*}{\textbf{Country}} & \multicolumn{5}{c}{\textbf{RMSE}} \\
    \cline{2-6}
    
    & \multicolumn{2}{c}{\textbf{LSTM}} & & \multicolumn{2}{c}{\textbf{Hybrid Model}} \\
    \cline{2-3}
    \cline{5-6}
    
    & \textbf{Case} & \textbf{Death} & & \textbf{Case} & \textbf{Death} \\
    \hline
    
     USA & $7934$ & $3135$ && $623$ & $1671$ \\
     ESP & $1512$ & $223$  && $342$ & $164$  \\
     ITA & $2729$ & $427$  && $207$ & $35$   \\
     FRA & $1846$ & $587$  && $424$ & $264$  \\
     DEU & $1070$ & $105$  && $173$ & $53$   \\
     GBR & $1509$ & $273$  && $735$ & $143$  \\
     TUR & $2924$ & $10$   && $101$ & $17$   \\
     IRN & $562$  & $32$   && $227$ & $20$   \\
     CHN & $223$  & $391$  && $77$  & $194$  \\
     RUS & $2273$ & $23$   && $153$ & $6$    \\
     AUS & $37$   & $3$    && $15$  & $1$    \\
     \hline
         
    \end{tabular}
    
    \label{tab:Comparison}
\end{table}

Figure \ref{fig:pic1} to Fig. \ref{fig:pic4} illustrates that the hybrid model was capable of more accurate prediction than the LSTM model with all case studies. Not only can the model predict the behavior/trend, it also provides accurate estimates of the numbers of cases and deaths under considerable uncertainty. In these figures, forecasts for the next 20 days are provided. As seen, Spain, France, Germany, China, and Australia are able to control the number of cases at zero cases per day. However, the numbers of cases in the USA and Turkey in the next 20 days are around 25,000 and 2,000 per day, respectively. Therefore, these countries need greater restrictions and improved public knowledge and behavior if they are to decrease the number of cases. During this period, there is insufficient control in Iran and the United Kingdom. As the model is sensitive to large fluctuations in cases and deaths, care should be taken to maintain a stable trend. Besides, Fig. \ref{fig:pic3} and Fig. \ref{fig:pic4} show that LSTM crashes when estimating cases in Australia and France because of negative values of cases. 

\section{Conclusion}\label{Conclusion}

A novel hybrid model using LSTM and dynamic behavior models was proposed to forecast the spread of Covid-19 in the most-affected countries. Many factors affect the spread of this virus, so it is very difficult to make the right predictions of cases and deaths. In this regard, the LSTM and dynamic behavioral models were used to model a dynamic system with a high level of fidelity. The results show that the hybrid model can accurately predict the spread of Covid-19 based on real data.

The proposed hybrid model provides robust estimates with the exclusive use of regional properties. Adding more modules and using real data for different modules can substantially improve the model’s performance. In this paper, a TM, FM (including medical staff and hospitals), and DM were constructed to forecast the spread of Covid-19 in the top ten most-affected countries. Public knowledge and behavior can directly impact the spread of Covid-19. Additionally, skilled medical staff and high-quality hospitals can control the outbreak.


\bibliographystyle{IEEEtran}
\bibliography{Refs}

\end{document}